\documentclass[letterpaper,titlepage,11pt]{article}

\usepackage{amssymb,amsmath,amsfonts,jheppub}
\usepackage{mathtools}
\usepackage{mathrsfs}
\usepackage{epsfig}
\usepackage[toc]{appendix}
\numberwithin{equation}{section}
\usepackage{graphicx}
\usepackage{subfig}
\usepackage{tikz}
\usepackage{tensor}
\usepackage{todonotes}
\usetikzlibrary{decorations.pathmorphing,calc}
\usepackage[english]{babel}

\usepackage{amsmath,amssymb}
\usepackage{verbatim}
\usepackage{graphicx}
\usepackage{color}

\DeclareFontFamily{OT1}{rsfs}{}
\DeclareFontShape{OT1}{rsfs}{m}{n}{ <-7> rsfs5 <7-10> rsfs7 <10->rsfs10}{}
\DeclareMathAlphabet{\mycal}{OT1}{rsfs}{m}{n}

\newcommand{\be}[1]{ \begin{equation}\label{#1}
}

\newcommand{\bea}[1]{\begin{eqnarray}\label{#1}
}
\newcommand{\eea}{
\end{eqnarray}}

\newcommand{\tr}{\textrm{tr}}

\newcommand{\eq}[2]{\begin{equation} #1 \label{#2} \end{equation}}
\newcommand{\eps}{\varepsilon}

\DeclareMathOperator{\extdm}{d}
\newcommand{\extd}{\extdm \!}

\newcommand{\vp}{\varphi}

\newcommand{\mytitle}{Soft hairy warped black hole entropy}

\title{\mytitle}

\subheader{TUW--17--07}

\newcommand{\dg}{\ast}
\newcommand{\ph}{\dagger}
\newcommand{\wm}{\ddagger}

\author[\dg]{Daniel Grumiller,}
\emailAdd{grumil@hep.itp.tuwien.ac.at}
\author[\ph]{Philip Hacker,}
\emailAdd{e1227036@student.tuwien.ac.at}
\author[\wm]{and Wout Merbis}
\emailAdd{wmerbis@ulb.ac.be}
\affiliation[\dg,\ph,\wm]{Institute for Theoretical Physics, TU Wien, Wiedner Hauptstr.~8-10/136, A-1040 Vienna, Austria}
\affiliation[\wm]{Universit\'e Libre de Bruxelles and International Solvay Institutes, Physique Th\'eorique et Math\'ematique,
Campus Plaine - CP 231, B-1050 Bruxelles, Belgium}

\abstract{
We reconsider warped black hole solutions in topologically massive gravity and find novel boundary conditions that allow for soft hairy excitations on the horizon. 
To compute the associated symmetry algebra we develop a general framework to compute asymptotic symmetries in any Chern--Simons-like theory of gravity. 
We use this to show that the near horizon symmetry algebra consists of two $\mathfrak{u}(1)$ current algebras and recover the surprisingly simple entropy formula $S=2\pi\,(J_0^+ + J_0^-)$, where $J_0^\pm$ are zero mode charges of the current algebras. 
This provides the first example of a locally non-maximally symmetric configuration exhibiting this entropy law and thus non-trivial evidence for its universality. 
}


\begin{document}

\maketitle

\section{Introduction}\label{se:1}

Recently a surprisingly simple entropy formula emerged in the near horizon description of three-dimensional black holes \cite{Afshar:2016wfy} and cosmologies \cite{Afshar:2016kjj}, 
\eq{
S=2\pi\,\big(J_0^+ + J_0^-\big) 
}{eq:i1}
where $J_0^\pm$ are zero modes of $\mathfrak{u}(1)$ current algebras associated with soft hairy boundary conditions.\footnote{%
The notion of ``soft hair'' was introduced by Hawking, Perry and Strominger \cite{Hawking:2016msc, Hawking:2016sgy} and refers to non-trivial zero energy excitations of black holes. The near horizon description that led to the result \eqref{eq:i1} was inspired by (but differs in essential details from) work by Donnay, Giribet, Gonz\'alez and Pino \cite{Donnay:2015abr, Donnay:2016ejv}.
} 
This entropy formula is universal in the sense that it applies to flat space and anti-de~Sitter (AdS), to Einstein gravity as well as theories with higher derivatives \cite{Setare:2016vhy} or higher spin \cite{Grumiller:2016kcp, Ammon:2017vwt}. However, so far all studies relied on locally maximally symmetric setups. 

To test universality of the entropy formula \eqref{eq:i1} beyond maximally symmetric cases we reconsider warped black holes \cite{Anninos:2008fx} in topologically massive gravity (TMG) \cite{Deser:1982vy, Deser:1982wh}. We impose novel boundary conditions for soft hairy warped black holes in line with the ones introduced in \cite{Afshar:2016wfy}. In order to derive the near horizon symmetry algebra from these boundary conditions we develop a general framework for deriving asymptotic symmetry algebras in any Chern--Simons-like (CS-like) theory of gravity \cite{Hohm:2012vh,Bergshoeff:2014bia,Merbis:2014vja}. These models include TMG, but are not restricted to this special case and hence our results are useful for a variety of theories of massive gravity in three dimensions  \cite{Bergshoeff:2009hq,Bergshoeff:2009aq,Bergshoeff:2013xma,Bergshoeff:2014pca,Afshar:2014ffa,Afshar:2014dta,Setare:2014zea,Adami:2017phg}. 

Using the general framework we are able to find the near horizon charges and their associated symmetry algebra. We find that even in this locally non-maximally symmetric case, the entropy of the warped black hole can be written in terms of the zero-modes of the near horizon $\mathfrak{u}(1)$ charges. This is further evidence for the conjecture that non-extremal black holes in three dimensions have an entropy that is universally given by the sum of zero modes of $\mathfrak{u}(1)$ current algebras.

This work is organized as follows. 
In section \ref{se:2} we summarize relevant aspects of warped black holes in TMG. 
In section \ref{se:3} we introduce new boundary conditions that allow soft hair excitations on warped black hole horizons.
In section \ref{se:3.4} we present general results for boundary charges and asymptotic symmetries in CS-like theories.
In section \ref{se:3too} we apply these results to deduce the warped near horizon charges and their algebra (including its central extensions).
In section \ref{se:4} we derive the entropy of soft hairy warped black holes and recover the universal result \eqref{eq:i1}.
In section \ref{se:5} we conclude. 
In the appendices we list our conventions (appendix \ref{app:A1}), provide explicit expressions for near horizon warped black hole solutions with soft hair in TMG (appendix \ref{app:A}), construct the warped black hole generalization of the boundary conditions of \cite{Donnay:2015abr,Donnay:2016ejv} (appendix \ref{app:A2}) and present a general CS-like derivation of the black hole entropy (appendix \ref{app:B}).

Before starting we mention some of our conventions here. We denote the Levi--Civita-symbol by $\epsilon$, with the sign convention $\epsilon_{tr\vp}=+1$, and the corresponding tensor by $\varepsilon$. The sign convention for the Ricci-tensor is fixed by $R_{\mu\nu} = +\partial_\lambda\Gamma^\lambda{}_{\mu\nu} - \dots$, and we use mostly plus signature.

\bigskip

{\em Note added:} While finishing our work we became aware of \cite{Adami:2017} which studies the near horizon geometry of warped black holes in generalized minimal massive gravity. The master thesis \cite{Evrard:2017} also considers near horizon symmetries for warped black holes and has the result \eqref{eq:i1} for entropy of warped black holes (albeit without soft hair).

\section{Warped black holes in topologically massive gravity}\label{se:2}

TMG (with negative cosmological constant $\Lambda=-1/\ell^2$) is a third-derivative action \cite{Deser:1982vy, Deser:1982wh},
\eq{
I_{\textrm{\tiny TMG}} = \frac{1}{16\pi G}\,\int\extd^3x\sqrt{-g}\,\Big(R + \frac{2}{\ell^2}\Big) +  \frac{1}{16\pi G}\,I_{\textrm{\tiny gCS}}
}{eq:shw1}
containing the gravitational Chern--Simons (CS) term
\eq{
I_{\textrm{\tiny gCS}} = \frac{1}{\mu}\,\int\Big(\frac12\,\Gamma\wedge\extd\Gamma + \frac13\,\Gamma\wedge\Gamma\wedge\Gamma\Big)
}{eq:shw3}
with the Christoffel connection one-form $\Gamma$. Here $G$ is Newton's constant and $\mu$ the CS coupling constant, which from now on we rescale by $3/\ell$ for later convenience
\eq{
\mu\ell = 3\nu\,.
}{eq:shw5}

Varying the action \eqref{eq:shw1} with respect to the metric leads to third-derivative equations of motion
\eq{
R_{\alpha\beta} + \frac{2}{\ell^2}\, g_{\alpha\beta} + \frac{\ell}{3\nu}\,C_{\alpha\beta} = 0 
}{eq:shw2}
where $C_{\alpha\beta}=\varepsilon_\alpha{}^{\gamma\delta}\nabla_\gamma(R_{\delta\beta}-\tfrac14\,g_{\delta\beta}R)$ is the Cotton tensor \cite{Garcia:2003bw}. We set $\ell=1$ from now on.

While every solution to three-dimensional Einstein gravity solves the TMG equations of motion \eqref{eq:shw2}, there are numerous other solutions, sometimes with striking asymptotic behavior (see e.g.~\cite{Ertl:2010dh} for a classification of all local solutions with two commuting Killing vectors and \cite{Chow:2009km} for further solutions). The set of solutions we are currently interested in are locally warped AdS solutions \cite{Vuorio:1985ta, Percacci:1986ja} with a black hole horizon \cite{Nutku:1993eb, Gurses:1994bjn, Bouchareb:2007yx, Clement:2009gq}. These black holes are quotients of warped AdS$_3$ \cite{Anninos:2008fx} in the same way as BTZ black holes \cite{Banados:1992wn} are quotients of AdS$_3$ \cite{Banados:1992gq} and solve the TMG equations of motion \eqref{eq:shw2} for $\nu>1$.

In ADM coordinates warped AdS black holes are given by the metric \cite{Anninos:2008fx}
\eq{
\extd s^2 = -N(\hat r)^2\, \extd \hat t^2 + \frac{\extd \hat r^2}{4R(\hat r)^2 N(\hat r)^2} + R(\hat r)^2\,\big(\extd\hat\vp+N^{\hat\vp}(\hat r)\extd \hat t\big)^2
}{eq:shw4}
with radial function $R(\hat r)$, lapse $N(\hat r)$ and shift $N^{\hat\vp}(\hat r)$ given by
\begin{align}
    R(\hat r)^2 &= \frac{\hat r}{4}\,\Big(3(\nu^2-1)\hat r+(\nu^2+3)(\hat r_+ + \hat r_-)-4\nu\sqrt{\hat r_+\hat r_-(\nu^2+3)}\Big) \\
    N(\hat r) &= \frac{\sqrt{(\nu^2+3)(\hat r-\hat r_+)(\hat r-\hat r_-)}}{2R(\hat r)} \\
    N^{\hat\vp}(\hat r) &= \frac{2\nu\hat r-\sqrt{\hat r_+ \hat r_-(\nu^2+3)}}{2R(\hat r)^2}\,. \label{eq:shw6}
\end{align}
The angular coordinate is periodic, $\hat\vp\sim\hat\vp+2\pi$, the radial coordinate is non-negative, $\hat r\geq 0$, and the time coordinate is unrestricted, $\hat t\in\mathbb{R}$. The locus $\hat r=\hat r_+$ corresponds to the black hole horizon, while $\hat r=\hat r_-$ is the inner horizon, where $\hat r_+>\hat r_->0$ are real parameters labelling all warped AdS black hole solutions. For later purposes we introduce new parameters $r_+>r_->0$ defined by
\eq{
r_+ = \hat r_+-\sqrt{\hat r_+\hat r_-}\qquad\qquad r_- = \sqrt{\hat r_+\hat r_-}-\hat r_- \,.
}{eq:angelinajolie}

\section{Soft hairy warped black holes}\label{se:3}

In this section we establish new boundary conditions for TMG that allow for ``warped black flowers'', i.e., warped AdS black holes equipped with arbitrary near horizon soft hair, in full analogy to the BTZ case studied originally \cite{Afshar:2016wfy}. In our whole analysis we assume $\nu>1$, but occasionally consider the BTZ limit $\nu\to 1$.

In section \ref{se:3.1} we recast the warped AdS black hole metric into a form suitable for a near horizon discussion. 
In section \ref{se:3.2} we propose new boundary conditions for warped AdS black holes that we call ``soft hairy boundary conditions''.
In section \ref{se:3.3} we present these boundary conditions in a convenient CS-like formulation.

\subsection{Near horizon line-element}\label{se:3.1}

The first step of our construction is to rewrite the warped AdS black hole metric \eqref{eq:shw4}-\eqref{eq:shw6} in a near horizon expansion as Rindler spacetime plus subleading terms.
\eq{
\extd s^2 = -a^2r^2\extd t^2 + \extd r^2 + \gamma^2\,\extd\vp^2+2a\omega r^2\,\extd t \extd\vp + \dots
}{eq:shw7}
The relations between hatted and unhatted coordinates are valid for $\hat r\geq\hat r_+$ and given by
\begin{align}
    t &= \hat t\\
    r &= \frac{2}{\sqrt{\nu^2+3}}\,\textrm{arcosh}\sqrt{(\hat r - \hat r_-)/(\hat r_+ - \hat r_-)}\\
    \vp &= \hat\vp + \frac{\hat t}{\nu\hat r_+-\tfrac12\,\sqrt{\hat r_+\hat r_-\,(\nu^2+3)}}\,.
\end{align}
Rindler acceleration $a$, horizon radius $\gamma$ and rotation parameter $\omega$ are determined from $r_\pm$.
\begin{align}
    a &= \frac{(\nu^2+3)(r_+^2-r_-^2)}{2r_+(2\nu r_+-\sqrt{\nu^2+3}\,r_-)} \\
    \gamma &= \frac{r_+(2\nu r_+-\sqrt{\nu^2+3}\,r_-)}{2(r_+-r_-)} \label{eq:lalapetz} \\
    \omega &= \frac{3 (1-\nu^2)\, r_+^2 + 2\nu\sqrt{\nu^2+3}\,r_+r_- -(\nu^2+3) r_-^2}{4(r_+-r_-)} \,.
    \label{eq:shw14}
\end{align}
Note the simple relation $r_+ + r_- = 4(\nu\gamma+\omega)/(\nu^2+3)$. 

For the special case $\nu=1$ the solutions describe BTZ black holes with outer/inner horizons at $\hat r=\hat r_\pm$. The relations above then simplify considerably.
\eq{
a = 1+\frac{r_-}{r_+} \qquad\qquad \gamma = r_+ \qquad\qquad \omega = r_- \,.
}{eq:shw12}
Note that the value for Rindler acceleration differs from the usual BTZ expression $a_{\textrm{\tiny BTZ}}=(r_+^2-r_-^2)/r_+$ since the time coordinate used in \eqref{eq:shw4} and \eqref{eq:shw7} differs by a factor $r_+-r_-$ from the usual BTZ time coordinate.

\subsection{Soft hairy boundary conditions in metric formulation}\label{se:3.2}

Concurrent with \cite{Afshar:2016wfy} we impose the following boundary conditions near the warped AdS black hole horizon $r=0$ (gauge fixed to Gaussian normal coordinates).
\begin{subequations}
\label{eq:shw8}
\begin{align}
    g_{tt} & = -a^2(t,\,\vp)\,r^2 + \Xi_{tt}(t,\,\vp)\, r^4 + {\cal O}(r^6) \\
    g_{t\vp} & = a(t,\,\vp)\,\omega(t,\,\vp)\, r^2  + \Xi_{t\vp}(t,\,\vp) r^4 + {\cal O}(r^6) \\
    g_{\vp\vp} & =\gamma^2(t,\,\vp) + \Xi_{\vp\vp}(t,\,\vp)\,r^2 + {\cal O}(r^4) \\
    g_{rr} & = 1 \qquad\qquad g_{tr} = g_{\vp r} = 0 
\end{align}
\end{subequations}
with the functions 
\begin{align}
\Xi_{tt}(t,\,\vp) &= a^2(t,\,\vp)\,(\tfrac23 \nu^2-1) \\ 
\Xi_{t\vp}(t,\,\vp)&=a(t,\,\vp)\,\big(\tfrac34 \nu (1-\nu^2) \gamma(t,\,\vp) + (1 - \tfrac23 \nu^2) \omega(t,\,\vp)\big) \\ 
\Xi_{\vp\vp}(t,\,\vp)&=\nu^2 \gamma(t,\,\vp)^2 - \omega(t,\,\vp)^2
\end{align}
chosen such that at $r=0$ the Ricci scalar and the square of the Ricci tensor are constants specific to warped AdS$_3$, $R=-6$ and $R_{\mu\nu}R^{\mu\nu}=6(3 - 2\nu^2 + \nu^4)$ with warping parameter $\nu > 1$. The leading order fluctuations of the metric are restricted by 
\eq{
\delta a(t,\,\vp) = 0\qquad\qquad \delta\gamma(t,\,\vp)\;\textrm{and\;} \delta \omega(t,\,\vp) \textrm{\;arbitrary}\,.
}{eq:shw9}
Note in particular that the function $a(t,\,\vp)$ (whose zero mode corresponds to Rindler acceleration) is not allowed to vary. Additional on-shell conditions are given by\footnote{%
Finiteness and constancy of the Ricci scalar at $r=0$ imply the differential equation $(\partial_t a)(\partial_t\gamma)=a\partial_t^2\gamma$, which is solved by $\gamma(t,\,\vp)=c_0(\vp)+\int^ta(t',\,\vp)c_1(\vp)\,\extd t'$. However, if $c_1\neq 0$ the function $\gamma$ diverges linearly in time for constant Rindler acceleration, which we consider as unphysical. Therefore, we set $c_1=0$.
}
\eq{
\partial_t\gamma(t,\,\vp) = 0 \qquad\qquad \partial_t\omega(t,\,\vp) = -\partial_\vp a(t,\,\vp) \,.
}{eq:shw11}
The boundary conditions \eqref{eq:shw8}-\eqref{eq:shw9} together with the on-shell conditions \eqref{eq:shw11} specify our theory and allow not only for all the warped AdS black hole solutions \eqref{eq:shw7}-\eqref{eq:shw14}, but additionally permit excitations on the horizon by allowing non-trivial functional dependence of $\gamma$ and $\omega$ on the coordinates. We shall recall in section \ref{se:3too} in which sense they can be interpreted as zero energy excitations, i.e., as soft hair. 

For simplicity we assume constant Rindler acceleration from now on, $a=\rm const$. The on-shell conditions \eqref{eq:shw11} then imply conservation equations that can be interpreted as ``near horizon holographic Ward identities'', analogous to the conservation equations of the \mbox{(anti-)}ho\-lo\-mor\-phic flux components of the stress-energy tensor in AdS$_3$/CFT$_2$. With this restriction, the class of metrics solving the TMG equations of motion \eqref{eq:shw2} for $\nu>1$ allowed by our boundary conditions \eqref{eq:shw8}-\eqref{eq:shw9} is given by
\begin{multline}
\extd s^2 = -a^2r^2\extd t^2 + \extd r^2 + \gamma^2(\vp)\,\extd\vp^2+2a\omega(\vp) r^2\,\extd t \extd\vp +  r^2\,\big(\nu^2 \gamma(\vp)^2 - \omega(\vp)^2\big)\,\extd\vp^2 \\
+ r^4\,\Big(\tfrac{a^2}{3}\,(2 \nu^2-3)\,\extd t^2 - \tfrac{a}{6}\big(9 \nu (\nu^2-1) \gamma(\vp) - 4 (3 - 2 \nu^2) \omega(\vp)\big)\,\extd t\extd\vp\\
+ \tfrac16\,(\nu\gamma + \omega)\big(\nu (5\nu^2-3)\,\gamma - 2 (3 - 2\nu^2)\,\omega\big) \,\extd\vp^2\Big)  + {\cal O}(r^6) \,.
\label{eq:shw15}
\end{multline}

The exact solution reproducing \eqref{eq:shw15} near the horizon is given by the line-element
\begin{multline}
\extd s^2 = -\,\big(a^2\rho^2-\tfrac34\,(\nu^2-1)\,a^2\rho^4\big)\,\extd t^2 + \frac{\extd\rho^2}{1+\tfrac14\,(\nu^2+3)\,\rho^2} \\
+ \big(\gamma(\vp)^2 +  (\nu^2\gamma(\vp)^2 - \omega(\vp)^2)\,\rho^2 + \tfrac34\,(\nu^2-1)\,(\nu\gamma(\vp) + \omega(\vp))^2\, \rho^4 \big)\,\extd\vp^2 \\
+2 \big(a\omega(\vp)\,\rho^2 - a(\nu\gamma(\vp) + \omega(\vp))\tfrac34\,(\nu^2 - 1)\,\rho^4\big)\extd t\extd\vp\,.
    \label{eq:shwBH}
\end{multline}
By analogy to the BTZ case \cite{Afshar:2016wfy, Afshar:2016kjj} we call metrics of the form \eqref{eq:shwBH} ``soft hairy warped black holes'' or ``warped black flowers''. The relation between the two radial coordinates is $\rho=2\sinh\big(\tfrac12\,\sqrt{\nu^2+3}\,r\big)/\sqrt{\nu^2+3}$, so that for small radii $\rho=r+\tfrac{1}{24}\,(\nu^2+3)\,r^3+{\cal O}(r^5)$. At large $\rho$ our boundary conditions can be contrasted to previous ones \cite{Compere:2009zj, Henneaux:2011hv}, where the leading order components in $g_{\vp\vp}$ and $g_{t\vp}$ are not allowed to vary, as opposed to here. This is similar to the corresponding comparison of near horizon and Brown--Henneaux boundary conditions for BTZ \cite{Afshar:2016wfy, Afshar:2016kjj}.

\subsection{Soft hairy boundary conditions in Chern--Simons-like formulation}\label{se:3.3}

Following \cite{Merbis:2014vja} we use a CS-like formulation to describe TMG and our new boundary conditions. The basic variables are the triad $e$, dualized spin-connection $\omega$ and the Schouten one-form $f$, all of which are $sl(2,\,\mathbb{R})$-valued one-forms. In terms of these, the bulk action is given by 
\eq{
I = -\frac{1}{4\pi G}\,\int\tr\Big(e\wedge\big(\extd\omega + \tfrac12\,[\omega\stackrel{\wedge}{,}\,\omega] + \tfrac{1}{6}\,[e\stackrel{\wedge}{,}\,e]\big) - \frac{1}{3\nu}\,f\wedge\big(\extd e+[\omega\stackrel{\wedge}{,}\,e]\big) - \frac{1}{6\nu}\omega\wedge\big(\extd\omega+\tfrac{1}{3}\,[\omega\stackrel{\wedge}{,}\,\omega]\big) \Big)\,.
}{eq:shw16}
Our conventions for the $sl(2,\,\mathbb{R})$ algebra and wedged commutators are collected in appendix \ref{app:A1}. The equations of motion are first order in derivatives and read
\begin{subequations}
\label{eq:EOM}
\begin{align}
    \extd e+ [\omega\stackrel{\wedge}{,}\,e] &=0\\
    \extd\omega + \tfrac12\,[\omega\stackrel{\wedge}{,}\,\omega] + [e\stackrel{\wedge}{,}\,f] &= 0\\
    \extd f+[\omega\stackrel{\wedge}{,}\,f]+3\nu\,[e\stackrel{\wedge}{,}\,f] -\frac{3\nu}{2}\,[e\stackrel{\wedge}{,}\,e] &= 0\,.
\end{align}
\end{subequations}

For constant Rindler acceleration our near horizon boundary conditions describing soft hairy warped black holes can be written succinctly as
\begin{subequations}
\label{eq:bcs}
\begin{align}
    e_t &= ar\,\frac{L_+ + L_-}{2}\,\big(1+{\cal O}(r^3)\big) \\
    e_\vp &= \gamma(\vp)\,\big(1 + \frac{\nu^2}{2}\,r^2\big) \,L_0 - r\,\omega(\vp)\,\frac{L_+ + L_-}{2} + {\cal O}(r^3) \\
    e_r &= \frac{L_+ - L_-}{2} \,.
\end{align}
\end{subequations}
Again we have partially gauge fixed to Gaussian normal coordinates with respect to the radial coordinate $r$. Additionally, we have gauge-fixed local Lorentz symmetries completely by demanding that $e_r$ is given by the combination $L_+-L_-$ and $e_t$ proportional to the combination $L_+ + L_-$. The warping parameter $\nu$ appears explicitly only in the $r^2$-term of the triad-component $e_\vp$ and implicitly in subleading terms. If $\nu=1$ the BTZ near horizon boundary conditions in the form presented in \cite{Afshar:2016kjj} are recovered. The expressions for the spin-connection $\omega$ and the Schouten one-form $f$ as well as the subleading terms in the triad not displayed above all follow from the equations of motion \eqref{eq:EOM} and are provided explicitly up to higher order terms in appendix \ref{app:A}. The metric given by \eqref{eq:shw17} coincides with \eqref{eq:shw15}. 

So far we have merely translated our results into first order form, but in the calculations in the next three sections this translation will pay off. Before we refocus on soft hairy warped black holes in TMG we present general results for boundary charges and asymptotic symmetries in CS-like theories in the next section.

\section{
Asymptotic symmetries in Chern--Simons-like theories}\label{se:3.4}

\newcommand{\ione}{{\tt p}}
\newcommand{\itwo}{{\tt q}}
\newcommand{\ithr}{{\tt r}}
\newcommand{\ifou}{{\tt s}}
\newcommand{\ifiv}{{\tt t}}

TMG falls into a larger class of theories characterized by a CS-like formulation \cite{Hohm:2012vh,Bergshoeff:2014bia}.  For future reference and convenience of application of these results to other theories of three-dimensional massive gravity we start with the more general CS-like formulation and restrict to TMG in later sections.

In section \ref{se:cs1} we summarize the Hamiltonian analysis of CS-like theories and identify the generator of diffeomorphisms.
In section \ref{se:cs2} we present general expressions for their canonical boundary charges and prove that the diffeomorphism generators indeed generate gauge symmetries.
In section \ref{se:cs3} we propose a systematic way to determine the asymptotic symmetries generated by the boundary charges. Here we work mostly in a Hamiltonian formulation of the theory, generalizing the work of \cite{Banados:1994tn} to CS-like models. Similar results have been obtained in \cite{Adami:2017phg} through the covariant phase space methods of \cite{Barnich:2001jy}.

\subsection{Hamiltonian analysis of CS-like theories}\label{se:cs1}

CS-like models can be defined in terms of a set of $sl(2,\,\mathbb{R})$-vector valued one-form fields labeled by field space indices\footnote{The field space indices label the fields by taking values equal to the symbol of the field. So in the case of TMG we denote the triad $e$ as $a^e$ and similarly $a^{\omega} =\omega$ and $a^f = f$.} $\ione,\itwo,\ithr,\ifou,\ifiv$, i.e., $a^\ione = a^n_\mu{}^\ione \, L_n\, \extd x^\mu$. The defining feature of these models is that they have a bulk action which is reminiscent of, but not quite equal to, the usual CS action in three dimensions
\eq{
    I = \frac{k}{2\pi} \int \tr \Big(g_{\ione\itwo}\, a^\ione \wedge\extd a^\itwo + \frac13\, f_{\ione\itwo\ithr}\, a^\ione\wedge a^\itwo \wedge a^\ithr \Big)\,.
}{eq:shw20}
Here $g_{\ione\itwo}$ and $f_{\ione\itwo\ithr}$ are a completely symmetric field space metric and structure constants, respectively and $k$ is the CS-like level, the overall coupling constant of the theory. For special values of the field space metric and structure constants the theory in fact is equal to a CS theory, but in general this is not the case. We assume that all of the fields appearing in the action have a kinetic term. This implies that $g_{\ione\itwo}$ is invertible and we use its inverse to raise field space indices, i.e. $f^\ione{}_{\itwo\ithr} = g^{\ione\ifou}f_{\ifou\itwo\ithr}$.

We consider now gauge-like transformations of the fields $a^\ione \to a^\ione + \delta_\xi a^\ione $ with
\eq{
    \delta_{\xi} a^\ione = \extd\xi^\ione + f^\ione{}_{\itwo\ithr}\,[a^\itwo,\,\xi^\ithr]\,.
}{eq:shw21}
Here $\xi^\ione = \xi^n{}^\ione L_n$ is an $sl(2,\mathbb{R})$ valued `field space' vector. Not all of these transformations correspond to gauge transformations since only for certain $\xi^\ione$ do they leave the action invariant. In the Hamiltonian formalism this statement means that not all corresponding constraints are necessarily first class. We will now show which constraints correspond to the generators of diffeomorphisms and prove that they are first class by computing their Poisson brackets with all other constraints explicitly in the next subsection. 

Analogously to the CS-formulation of Einstein gravity \cite{Achucarro:1986vz, Witten:1988hc} we find that diffeomorphisms are generated by the transformations \eqref{eq:shw21} with $\xi^\ione$ chosen as
\eq{
    \xi^\ione = a_\nu{}^\ione \zeta^\nu\,.
}{eq:diffeos}
With this choice of parameters \eqref{eq:shw21} can be written as
\eq{
    \delta_{\zeta} a_{\mu}{}^\ione = \zeta^{\nu} \partial_{\mu}a_{\nu}{}^\ione + a_{\nu}{}^\ione \partial_{\mu} \zeta^{\nu} + \ldots \stackrel{{\rm on-shell}}{=} {\cal L}_{\zeta} a_{\mu}{}^\ione
}{eq:CSlikediffeo}
where the ellipsis refers to terms which vanish by use of the field equations
\eq{
    g_{\ione\itwo}\, \extd a^\itwo + \frac12\, f_{\ione\itwo\ithr}\, [a^\itwo \stackrel{\wedge}{,}\, a^\ithr] = 0\,.
}{eq:CSlikeeom}

In \cite{Bergshoeff:2014bia} the Hamiltonian form of CS-like theories was studied. There it was shown that the Hamiltonian solely consists of primary constraints $\phi_\ione$ times Lagrange multipliers $a_t{}^\ione$, which are the time components of the fields. 
\eq{
    {\cal H} = - \int \extd^2x\,\tr(  a_{t}{}^\ione\, \phi_\ione) \equiv -\frac{k}{\pi} \int \extd^2x \, \tr\Big( a_t{}^\ione\, \epsilon^{ij} \big( g_{\ione\itwo}\, \partial_i a_j{}^\itwo + \frac12\, f_{\ione\itwo\ithr}\, [a_i{}^\itwo,\,a_j{}^\ithr] \big) \Big)
}{eq:CSlikeH}
Here the integration is over the spatial part of the manifold, which we will denote with the two-dimensional Latin indices $i,j$, and $\epsilon^{ij} \equiv \epsilon^{tij}$.  In our conventions the Poisson brackets of the canonical fields are 
\begin{equation}\label{poissonbr}
\left\{a_{i}^n{}^\ione(x),\, a_{j}^m{}^\itwo(y)\right\} = \frac{\pi}{k}\, \epsilon_{ij}\, g^{\ione\itwo}\, \gamma^{nm}\, \delta^{(2)} (x-y)\,.
\end{equation}
where $\gamma^{nm}$ is the inverse of the $sl(2,\mathbb{R})$ invariant Killing form $\gamma_{nm}= \text{antidiag}(-1,\tfrac12,-1)_{nm}$.

The Poisson brackets of the constraint functions $\phi_\ione$, when integrated against the parameter $\xi^\ione$, generate the spatial components of the transformations \eqref{eq:shw21}.
\eq{
    \{ \phi[\xi^\itwo],a_i{}^\ione(y) \} \equiv  \left\{ \frac{k}{\pi}\int d^2x\; \tr\left( \xi^\itwo(x) \phi_\itwo(x) \right),\, a_i{}^\ione(y)\right\} = \delta_\xi a_{i}{}^\ione(y)
}{eq:trafo}
The time components of the fields are either fixed by consistency of time evolution of the constraints or left arbitrary if the associated constraint is first class and generates gauge transformations. If the field space metric and structure constants are such that all constraints are first class, then the associated CS-like theory is actually a CS theory of the Lie algebra spanned by the Poisson brackets of the constraints. In the most general case, and also for TMG, this is not true and not all of the constraints are first class. The precise conditions under which secondary and second class constraints arise in CS-like models was studied in detail in \cite{Bergshoeff:2014bia}. For this work it suffices to use the intuition from gravity and focus on constraints which generate diffeomorphisms, i.e. have gauge parameter which can be written as \eqref{eq:diffeos}. In that case, one can show that the constraint function $\phi[\xi^\ione]$ defined above generates diffeomorphisms.
\eq{
    \big\{ \phi[\xi^\itwo=a_{\mu}{}^\itwo\zeta^{\mu}],\,a_i{}^\ione(y) \big\}  =  {\cal L}_{\zeta} a_{i}{}^\ione(y)
}{eq:trafotoo}
This expression is true off-shell. Taking the symmetry parameter $\xi^r$ proportional to the fields generates an extra term in the Poisson brackets proportional to the constraints $\phi_\itwo$ which exactly cancels the ellipsis in \eqref{eq:CSlikediffeo}.

\subsection{Boundary charges in CS-like theories}\label{se:cs2}

In order to make the constraint functions well-defined on manifolds with a boundary we need to add a boundary term $Q[\xi^\ione]$ to them (see \cite{Regge:1974zd})
\begin{equation}\label{eq:Phidef}
\Phi[\xi^\ione] = \frac{k}{\pi}\,\int\extd^2 x \; \tr\big(\xi^\ione(x) \phi_\ione (x) \big) + Q[\xi^\ione] \,.
\end{equation}
This term is defined such that it cancels the boundary terms coming from the variation of the constraints with respect to the fields,
\eq{
    \delta Q[\xi^\ione] = - \frac{k}{\pi}\, \oint \tr \big(g_{\ione\itwo}\, \xi^\ione\, \delta a_\vp^\itwo\big)\extd\vp \,.
}{eq:shw22}
Here we assumed the boundary of the spatial manifold to be a circle parametrized by the coordinate $\vp$. One can now choose boundary conditions such that this expression is integrable, finite and conserved. 

Including this boundary term the Poisson brackets of the constraints (for general $\xi^\ione$ and $\eta^\itwo$) are \cite{Bergshoeff:2014bia}
\begin{subequations}
\label{eq:gen_poissonbr}
\begin{align} 
& \left\{  \Phi[\xi^\ione],\, \Phi[\eta^\itwo]  \right\} = \; \frac{k}{\pi}\, \int\extd^2 x \; \Big[\tr(f_{\ione\itwo}{}^\ithr [\xi^\ione,\,\eta^\itwo] \phi_\ithr)  \\
&\qquad + 
\epsilon^{ij} \left(f^\ithr{}_{\ione[\ifiv} f_{\ifou]\itwo\ithr} \tr(\xi^\ione\eta^\itwo)  \tr( a_i{}^\ifou a_j{}^\ifiv)  +  2f^\ithr{}_{\ione[\itwo} f_{\ifiv]\ifou\ithr} \tr(\xi^\ione a_i{}^\ifou)\tr(\eta^\itwo a_j{}^\ifiv)  \right)\Big] \label{eq:disaster}
\\ &\qquad + \frac{k}{\pi}\, \oint \extd\vp \; \tr \Big( \xi^\ione \big(g_{\ione\itwo}  \partial_\vp \eta^\itwo + f_{\ione\itwo\ithr} [a_{\vp}{}^\itwo,\, \eta^\ithr]   \big) \Big) 
\end{align}
\end{subequations}
In order for a constraint to be first class, its Poisson brackets with all other constraints should vanish weakly (on-shell). Hence the bulk part of the right hand side of the above expression should vanish up to terms proportional to the constraints themselves. The first term in \eqref{eq:gen_poissonbr} is already proportional to the constraints, but it is not yet improved by a boundary term. The last term is a boundary term, which should equal the charge needed to improve the first term, plus a possible central extension. The term in the middle is more troublesome and does not vanish for general $\xi^\ione$. Fortunately, when the parameter $\xi^\ione$ is chosen as in \eqref{eq:diffeos} this term vanishes. To show this rather non-trivial looking result, one should make use of the following identity which holds on-shell and can be derived by acting on \eqref{eq:CSlikeeom} with an exterior derivative and using the field equations once more. 
\eq{
f^\ithr{}_{\ione[\ifiv}f_{\ifou]\itwo\ithr} a^\ione \wedge \tr( a^\ifou \wedge a^\ifiv) = 0 
=\epsilon^{ij}\left( f^\ithr{}_{\ione[\ifiv}f_{\ifou]\itwo\ithr} a_t{}^\ione \tr(a_i{}^\ifou a_j{}^\ifiv)+ 2 f^\ithr{}_{\ione[\itwo} f_{\ifiv]\ifou\ithr} \tr(a_t{}^\ione a_i{}^\ifou) a_j{}^\ifiv \right) 
}{Intcon}
The terms that remain after plugging the identity \eqref{Intcon} into \eqref{eq:disaster} can be written as an anti-symmetric combination of three two-dimensional indices
\eq{
    \epsilon^{ij} \zeta^k f^\ithr{}_{\ione[\ifiv}f_{\ifou]\itwo\ithr} \tr(\eta^\itwo a_{[k}{}^\ione) \tr(a_i{}^\ifou a_{j]}{}^\ifiv)
}{eq:zero}
and hence \eqref{eq:disaster} vanishes.

The last thing we need to show is that the Poisson brackets of $\Phi[\xi^\ione=a_{\mu}{}^\ione\zeta^{\mu}]$ with any possible secondary constraints also weakly vanishes. In \cite{Bergshoeff:2014bia} it was shown that assuming (some of) the one-form fields $a^\ione$ to be invertible can lead to secondary constraints. This happens when the inverse of this field can be used to turn the three-form equation \eqref{Intcon} into a two-form identity. The spatial components of this two-form then constitute a secondary constraint on the canonical variables. In general these secondary constraints, labelled by $I$ take the form
\eq{
    \Psi_{I} = h_{I,\,\ione\itwo}\, \epsilon^{ij}\, \tr\big(a_{i}{}^\ione a_{j}{}^\itwo\big)
}{eq:seccon}
for some anti-symmetric field space matrix $h_{I,\,\ione\itwo}$. The Poisson bracket of the diffeomorphism generator $\Phi[\xi^\ione=a_{\mu}{}^\ione\zeta^{\mu}]$ with these constraints is
\eq{
    \{\Phi[\xi^\ione=a_{\mu}{}^\ione\zeta^{\mu}],\Psi_I \} = \zeta^{\mu}\partial_{\mu} \Psi_I\,.
}{eq:pbseccon}
Now we have shown that the constraints \eqref{eq:Phidef} with gauge parameter \eqref{eq:diffeos} generate diffeomorphisms and have weakly vanishing Poisson brackets with themselves and all the other constraints in the theory\footnote{In principle one would have to check for further ternary constraints. However, under the assumptions listed in \cite{Bergshoeff:2014bia} and in all the models of interest that we know of (including TMG) there are none.}. Hence these constraints are first class in the general CS-like model and thus generate gauge symmetries. Their boundary terms \eqref{eq:shw22} constitute the boundary charges. 

Now we proceed to impose suitable boundary conditions. Suitable in this case means strict enough to make the boundary charges \eqref{eq:shw22} integrable, finite and conserved, but loose enough to allow for a non-trivial asymptotic symmetry algebra generated by these boundary charges.   

\subsection{Asymptotic symmetry algebra in CS-like theories}\label{se:cs3}

In order to find the asymptotic symmetry algebra in the general CS-like theories we proceed in the following way. First we specify the boundary conditions for our fields $a^\ione$. They have to solve the constraint equations (at least asymptotically) and they should come equipped with the specification of what is allowed to fluctuate on the boundary and what is kept fixed, i.e., which components of the fields carry state-dependent information. 

Then we determine the transformations \eqref{eq:shw21} with gauge parameter \eqref{eq:diffeos} that preserve the boundary conditions, up to the transformation of state-dependent functions. In other words, on the left hand side of \eqref{eq:shw21} we specify which components of the fields are allowed to fluctuate. Then we find the asymptotic gauge parameters $\xi^\ione$ by solving for the right hand side of \eqref{eq:shw21}, usually (but not necessarily) in some asymptotic (or near horizon) expansion in the radial coordinate. 

In this process we can be assisted by the secondary constraints that follow from \eqref{Intcon}. 
\eq{
    h_{I,\ione\itwo}\tr(a^\ione \wedge a^\itwo) = 0
}{eq:secconCSlike}
Using this equation and \eqref{eq:diffeos} we find that
\eq{
    h_{I,\ione\itwo} \tr(a^\ione \; \xi^\itwo) = h_{I,\ione\itwo} \tr(a^\itwo\; \xi^\ione) 
}{eq:xiID}
Depending on the actual form of the secondary constraints, this could turn into a simple equation relating different components of $\xi^\ione$ to each other.

After having found the gauge parameters which preserve \eqref{eq:shw21}, the consistency of the boundary conditions can be checked by inserting the result for the gauge parameter into the variation of the charges \eqref{eq:shw22}. This should be finite on the boundary, integrable and conserved. Once these conditions are met, the Poisson brackets \eqref{eq:gen_poissonbr} will solely receive contributions from the boundary charges on-shell and reduce to the Dirac bracket algebra of boundary charges
\eq{
    \{ Q[\xi^\ione],Q[\eta^\itwo] \}^* = - \delta_{\eta} Q[\xi^\ione] =  \frac{k}{\pi}\, \oint \extd\vp \; \tr\left(g_{\ione\itwo}\, \xi^\ione \,\delta_{\eta} a_{\vp}{}^\itwo \right)
}{eq:shw23}
Imposing boundary conditions on $a_{\vp}{}^\ione$ suffices to determine the asymptotic symmetry algebra. The conditions on the radial component of the fields can be derived by solving the constraints asymptotically. This is often simplified by choosing a suitable local Lorentz gauge. The time components of the fields can then be found by demanding the boundary conditions on $a_{\vp}{}^\itwo$ to be conserved under time evolution. This condition amounts to
\eq{
    \partial_t a_{\vp}{}^\itwo = - \{{\cal H},\,a_{\vp}{}^\itwo \} = \{ \phi[a_t{}^\ifou],\, a_{\vp}{}^\itwo \} = \delta_{\xi^\ifou = a_t{}^\ifou} a_{\vp}{}^\itwo
}{eq:timecomp}
and implies that the time component of the fields can be taken to be equal to the transformations which preserve $a_{\vp}{}^\itwo$. Then the transformation of the state-dependent functions (which constitute the boundary Ward identities) will turn into bulk equations of motion. Any free functions appearing in the gauge parameter $\xi^\ione$ will become the theories chemical potentials. This situation is essentially equivalent to that of actual CS theories  \cite{Henneaux:2013dra}.

In practice, however, one often deals with CS-like theories of gravity where one of the one-forms is a triad. It is then natural to impose boundary conditions on the triad components inspired by some asymptotic (or near horizon) expansion of a metric in a second order formulation of the theory. The process of adding chemical potentials as described above could then be used to generalize a specific set of boundary conditions and find the corresponding metric with arbitrary chemical potentials switched on.

Now that we have discussed in detail the procedure to obtain the asymptotic symmetry algebra of boundary diffeomorphism charges in the general CS-like model \eqref{eq:shw20}, we proceed to apply this to the case of near horizon warped black hole boundary conditions in TMG.

\section{Warped near horizon symmetries}\label{se:3too}

As next step in our analysis we apply the results of the previous section to our boundary conditions \eqref{eq:bcs} in TMG. We start by finding the boundary condition preserving transformations in section \ref{se:3.new}. Then we derive the warped near horizon charges in section \ref{se:3.5}, showing that they are finite, integrable and conserved in time. In section \ref{se:3.6} we obtain the near horizon symmetry algebra associated with the warped near horizon charges.

\subsection{Boundary condition preserving transformations}\label{se:3.new}

In the case of TMG we recover from \eqref{eq:shw20} the bulk action \eqref{eq:shw16} by taking the CS-like level $k = \frac{1}{4G}$, and using the field space metric and structure constants
\begin{subequations}
\label{eq:shw24}
\begin{align}
    g_{e\omega} & = -1 & g_{\omega\omega} & =  \frac{1}{3\nu} & g_{ef} &  = \frac{1}{3\nu} \\
    f_{e\omega\omega} & = -1 & f_{eee} & = -1 & f_{e\omega f} & =  \frac{1}{3\nu} & f_{\omega\omega\omega} & =  \frac{1}{3\nu} \,.
\end{align}
\end{subequations}
The triad, spin-connection and Schouten one-form transform as \eqref{eq:shw21}, which in this case reads
\begin{subequations}
\label{eq:shw25}
\begin{align}
    \delta_\xi e & = \extd\xi^e + [\omega,\,\xi^e] + [e,\,\xi^{\omega}] \\
    \delta_\xi \omega & = \extd\xi^{\omega} + [\omega,\,\xi^{\omega}] + [e,\,\xi^{f}] + [f,\,\xi^e] \\
    \delta_\xi f & = \extd\xi^f + [\omega,\,\xi^f] + [f,\,\xi^{\omega}] + 3\nu \big( [e,\,\xi^f]+[f,\,\xi^e]-[e,\,\xi^e] \big)\,.
\end{align}
\end{subequations}
We are looking for the parameters $\xi^\ione$ which preserve the near horizon boundary conditions \eqref{eq:bcs} in the limit as $r\to 0$. To this end we can use the field equations to solve asymptotically for $\omega$ and $f$ (see appendix \ref{app:A}). The state-dependent functions in this solution are $\gamma(\vp)$ and $\omega(\vp)$ and hence we solve for $\xi^\ione$ that satisfy
\begin{subequations}
\label{eq:shw26}
\begin{align}
    \delta_\xi e_\vp &= \delta \gamma(\vp) \left(1 + {\cal O}(r^2)  \right)L_0 - \delta \omega(\vp) (r + {\cal O}(r^3) )\frac{L_+ + L_-}{2} \displaybreak[1]\\
    \delta_\xi \omega_{\vp} &= \delta \omega(\vp)(-1 + {\cal O}(r^2))L_0 + \delta\gamma(\vp)(\nu^2 r + {\cal O}(r^3) )\frac{L_+ + L_-}{2} \displaybreak[1]\\
    \delta_\xi f_{\vp} &= \delta \gamma(\vp)\left(\tfrac12(4\nu^2-3) + {\cal O}(r^2)  \right)L_0  \nonumber \\
    &\quad +  \left( \delta \gamma(\vp) \left(3\nu (\nu^2-1) -\delta\omega(\vp)\,(\tfrac32-\nu^2)\,\right)\, r + {\cal O}(r^3)  \right)\,\frac{(L_+ + L_-)}{2}
\end{align}
\end{subequations}
while all other components of variations of the fields must vanish. We are aided in this process by the secondary constraint of TMG, which reads (in covariant form)
\eq{
    \tr(e \wedge f) = 0\,.
}{eq:TMGseccon}
In that case \eqref{eq:xiID} turns into
\eq{
    \tr(e\, \xi^f) = \tr(f\, \xi^e)\,,
}{eq:xiIDTMG}
which we can solve for $\xi^f$ by invertibility of the triad.

The other components of $\xi^\ione$ are found by solving the conditions \eqref{eq:shw26}. It turns out that they can be written in terms of two arbitrary functions of $\vp$, which we call $\eta(\vp)$ and $\eps(\vp)$.
\begin{subequations}
\label{eq:shw27}
\begin{align}
    \xi^e = & \,-  \eps(\vp)\, L_0 + \eta(\vp)\, r\, \frac{L_+ + L_-}{2} + {\cal O}(r^2) \displaybreak[1]\\
    \xi^\omega = &\,  \eta(\vp)\, L_0 - \nu^2\, \eps(\vp)\, r\,  \frac{L_+ + L_-}{2}  + {\cal O}(r^2) \displaybreak[1]\\
    \xi^f =  & \, \tfrac12(4\nu^2-3)\eps(\vp)\,L_0 - 
  \left( 3\nu (\nu^2-1)\eps(\vp) - (\tfrac32-\nu^2)\eta(\vp)\,\right)\, r \,\frac{(L_+ + L_-)}{2}  + {\cal O}(r^2)
\end{align}
\end{subequations}
These transformations correspond to diffeomorphisms by the near horizon Killing vector $\zeta = \zeta^{\mu}\partial_\mu$ 
\eq{
    \zeta = \frac{\gamma(\vp)\eta(\vp) - \omega(\vp) \eps(\vp)}{a\gamma(\vp)}\partial_t - \frac{\eps(\vp)}{\gamma(\vp)}\partial_\vp + {\cal O}(r^2)
}{eq:askilvec}
The leading order of this Killing vector is exactly the same as in \cite{Afshar:2016kjj} [Eq.~(84) with $\Omega=0$]. 

In accordance with the near horizon boundary conditions for non-extremal black holes in Einstein gravity \cite{Afshar:2016wfy} we are interested here in $\xi^\ione$ that do not depend on the functions characterizing the specific state, $\gamma(\vp)$ and $\omega(\vp)$. As a result the near horizon Killing vector \eqref{eq:askilvec} does depend on these functions. In appendix \ref{app:A2} we explore the consequences of taking the Killing vector to be state-independent. This leads to a generalization of the work of \cite{Donnay:2015abr} to near horizon warped black holes.

We should also note here that the form of \eqref{eq:shw27} is exactly that of $a_t{}^\ione$ for $\ione= e, \omega, f$ given in appendix \ref{app:A}, if we identify the zero-mode of $\eta(\vp)$ with the Rindler acceleration $a$ and set $\eps(\vp)$ to zero. This suggests that one could allow for a second chemical potential in the metric by allowing for non-zero $\eps(\vp)$ in $a_t{}^\ione$. We have excluded this possibility from the metric perspective, as this would lead to ${\cal O}(r^0)$ terms in $g_{tt}$ and hence spoil the regularity of the solution on the horizon.

\subsection{Warped near horizon charges}\label{se:3.5}

Under the boundary condition preserving transformations \eqref{eq:shw25} generated by \eqref{eq:shw27} the state dependent functions $\gamma(\vp)$ and $\omega(\vp)$ transform as
\eq{
    \delta \gamma(\vp) = - \partial_{\vp}\eps \qquad \qquad \delta \omega(\vp) =  - \partial_{\vp} \eta\,.
}{eq:shw28}
The variation of the charges is readily computed through \eqref{eq:shw22} and easily integrated to
\eq{
    Q[\eps,\,\eta] = \frac{1}{8\pi G}  \oint \extd \vp\, \Big[ \eta(\vp)  \Big( \gamma(\vp) + \frac{\omega(\vp)}{3\nu} \Big) + \eps(\vp)\Big(\omega(\vp) + \frac{4\nu^2-3}{3\nu}\,\gamma(\vp) \Big) \Big]\,.
}{eq:Qwarp}
This is one of our main results. The charges \eqref{eq:Qwarp} are non-trivial, finite, integrable and conserved in time, $\partial_t Q[\eps,\,\eta]=0$. This proves that our starting point, the boundary conditions \eqref{eq:bcs}, was meaningful.

The boundary Hamiltonian, i.e., the charge associated with unit time-translations is given by
\eq{
H = Q\big|_{\zeta=\partial_t} = Q[0,\,a] = a\,\frac{k}{2\pi}\, \oint \extd \vp\, \Big( \gamma(\vp) + \frac{\omega(\vp)}{3\nu} \Big) \,.
}{eq:H}

\subsection{Warped near horizon symmetry algebra}\label{se:3.6}

The symmetry algebra \eqref{eq:shw23} of the warped near horizon charges \eqref{eq:Qwarp} with variations given by \eqref{eq:shw28} is conveniently expressed in terms of Fourier modes $J_n$ and $K_n$ defined by
\eq{
    J_n \equiv Q[\eps = 0, \eta = e^{in\vp}]\qquad \qquad K_n \equiv Q[\eps = e^{in\vp}, \eta = 0]\,.
}{eq:shw29}
These modes satisfy the commutation relations
\begin{subequations}
\label{eq:nhalg}
\begin{align}
    [J_n,\,J_m] & = \frac{k}{3\nu}\, n\, \delta_{n+m,\,0} \\
    [J_n,\,K_m] & = k\, n\, \delta_{n+m,\,0} \\
    [K_n,\,K_m] & = \frac{k(4\nu^2 - 3)}{3\nu}\, n\, \delta_{n+m,\,0}\,.
\end{align}
\end{subequations}
For finite $\nu\geq 1$ the algebra \eqref{eq:nhalg} can always be diagonalized 
\eq{
J_n^\pm = \frac12\,\Big(J_n \pm \frac{1}{\sqrt{4\nu^2-3}}\,K_n\Big)
}{eq:shw31}
leading to our final result for the warped near horizon symmetry algebra, the non-vanishing commutators of which read
\eq{
 [J_n^\pm,\,J_m^\pm] = \pm \frac{k^\pm_{(\nu)}}{2}\,n\,\delta_{n+m,\,0}
}{eq:shw32}
with the left- and right-$\mathfrak{u}(1)$ levels
\eq{
k^\pm_{(\nu)} = \frac{k}{\sqrt{4\nu^2-3}}\,\Big(1\pm\frac{\sqrt{4\nu^2-3}}{3\nu}\Big)\,.
}{eq:shw33}

The limiting case $\nu=1$ describes BTZ black holes in TMG and yields $\mathfrak{u}(1)$ levels $k^+_{(1)}=\tfrac43\,k$, $k^-_{(1)}=\tfrac23\,k$, so that we get
\eq{
k^+_{(1)} + k^-_{(1)} = 2k\qquad k^+_{(1)} - k^-_{(1)} = \frac23\,k\,.
}{eq:shw34}
Comparing with the known results for the left and right central charges of TMG \cite{Kraus:2005zm, Kraus:2006wn}
\eq{
c^+_{(1)} + c_{(1)}^- = 12k\qquad c^+_{(1)} - c^-_{(1)} = 4k
}{eq:shw35}
shows consistency with \eqref{eq:shw34} provided we make the usual identification between central charges and levels, $c_{(1)}^\pm = 6\,k_{(1)}^\pm$. Thus, as expected the difference between the $\mathfrak{u}(1)$ levels $k^\pm_{(1)}$ is a measure for the gravitational anomaly, while their sum is a measure for the conformal anomaly. 

Thus, we have extended one of the main results of previous near horizon analyses \cite{Afshar:2016wfy, Afshar:2016kjj} to locally non-maximally symmetric solutions. Namely, the near horizon symmetry algebra consists of two $\mathfrak{u}(1)$ current algebras or, equivalently, of infinitely many Heisenberg algebras together with two zero mode charges $J_0^\pm$. The discussion of soft hair excitations in these papers generalizes straightforwardly to the present case. This is so, because the near horizon Hamiltonian $H$ \eqref{eq:H} is a sum of zero mode charges 
\eq{
H = a\,\big(J_0^+ + J_0^-\big) = a\,J_0
}{eq:nhH}
and therefore commutes with all raising operators $J_{-n}^\pm$ so that, like in previous cases, all soft hair descendants have the same energy as the parent state.

\section{Entropy}\label{se:4}

Thermodynamics of warped black holes in TMG was studied in \cite{Anninos:2008fx}, where they found a macroscopic result for the entropy
\eq{
S=\frac{\pi}{24\nu G}\,\Big((9\nu^2+3)\,\hat r_+ -(\nu^2+3)\,\hat r_- - 4\nu\sqrt{\hat r_+ \hat r_-(\nu^2+3)}\Big)
}{eq:SALPSS}
and suggested a microscopic one of Cardy-type that matches the result above. Given that the asymptotic symmetries of warped black holes are of warped CFT type \cite{Compere:2008cv} a slightly more natural microscopic formula for these black holes that also matches the macroscopic result \eqref{eq:SALPSS} was proposed in \cite{Detournay:2012pc} based on a Cardy-like formula for warped CFTs. 

The conjecture that we want to test in the current paper is whether or not the macroscopic entropy \eqref{eq:SALPSS} again has the simple form in terms of near horizon variables given by \eqref{eq:i1}. The algebraic relations \eqref{eq:lalapetz} and \eqref{eq:shw14} allow to express the entropy \eqref{eq:SALPSS} as
\eq{
    S = \frac{2\pi}{4G}\, \Big( \gamma + \frac{\omega}{3\nu} \Big) = 2\pi\, J_0 \,.
}{eq:entropy1}
In the second equality of the simple result \eqref{eq:entropy1} we used the definition of the near horizon charges \eqref{eq:Qwarp} in terms of Fourier modes \eqref{eq:shw29}. Comparing the result for entropy \eqref{eq:entropy1} with the near horizon Hamiltonian \eqref{eq:nhH} and using the standard relation between Rindler acceleration and Unruh temperature, $a=2\pi T$, yields $H=TS$, so that we recover the expected near horizon first law \cite{Halyo:2014nha, Donnay:2015abr, Afshar:2016wfy}
\eq{
\extd H = T \,\extd S\,.
}{eq:1stlaw}

The final step is to use the diagonal basis \eqref{eq:shw31} to bring the entropy into the form conjectured in the introduction
\eq{
S  = 2\pi\,\big(J_0^+ + J_0^-\big)\,.
}{eq:entropy}
In appendix \ref{app:B} we derive a general result for the entropy in the CS-like formulation, which reproduces the entropy \eqref{eq:entropy} for soft hairy warped black holes in TMG.

Our main result \eqref{eq:entropy} shows that the entropy of warped black holes (and their soft hairy generalizations constructed in the present work) is the sum of zero mode charges of two commuting $\mathfrak{u}(1)$ current algebras that arise in their near horizon description.

\section{Conclusions}\label{se:5}

We provided the first example of locally non-maximally symmetric black holes that exhibit the entropy law \eqref{eq:entropy}, conjectured to be universal in \cite{Afshar:2016kjj}, which provides highly non-trivial evidence for the conjecture. While doing so, we have established novel boundary conditions \eqref{eq:shw8}-\eqref{eq:shw9} [or, equivalently, \eqref{eq:bcs} in first order formulation for constant Rindler acceleration] that allow soft hair excitations of warped black holes in TMG.

Our example is the first of possibly numerous others. There are two natural generalizations. One could consider soft hairy warped black holes in other higher derivative/CS-like theories of three-dimensional gravity (see e.g.~\cite{Merbis:2014vja} and refs.~therein), and/or one could study other locally non-maximally symmetric black hole solutions (see e.g.~\cite{Chow:2009km} and refs.~therein for such solutions in TMG). To this end our general framework for computing canonical boundary charges and asymptotic symmetry algebras in Chern--Simons-like theories of gravity will surely be useful.

It would be very interesting to verify if the ``fluffball'' proposal for semi-classical near horizon microstates \cite{Afshar:2016uax, Sheikh-Jabbari:2016npa, Afshar:2017okz} works also for warped black holes. Relatedly (but also independently) it would be gratifying to know the precise coefficient of the log corrections to (soft hairy) warped black holes. We expect the calculation to be analogous to the BTZ case \cite{Grumiller:2017jft}, with the only subtlety being the use of an ensemble different from the usual ones that arise in asymptotic discussions (see e.g.~\cite{Sen:2012dw} regarding the role played by the choice of thermodynamical ensemble for the numerical coefficient in the log-corrections to black hole entropy).

Finally, it could be rewarding to lift our results to dimensions higher than three. There are already indications that a similar entropy law exists in four spacetime dimensions \cite{Gonzalez:2017sfq} based on four $\mathfrak{u}(1)$ current algebras \cite{Afshar:2016uax} and that the ``fluffball'' proposal could generalize as well \cite{Hajian:2017mrf}, but it remains to be seen how universal are the higher-dimensional generalizations of the entropy formula \eqref{eq:i1}.

We conclude with restating the initial conjecture with more confidence: non-extremal black holes in three dimensions (in Einstein gravity, higher derivative and/or higher spin gravity) have an entropy of the form \eqref{eq:i1}, where $J_0^\pm$ are zero modes of $\mathfrak{u}(1)$ current algebras that arise as near horizon symmetries. It would be excellent to find a generic proof of this conjecture or a non-trivial counter-example.\footnote{%
There is a somewhat trivial counter-example. If one considers black holes (or flat space cosmologies) in three-dimensional higher spin theory that are not continuously connected to the pure spin-2 branch then the entropy can also depend on additional $\mathfrak{u}(1)$ zero mode charges and not just on the two zero-mode charges associated with spin-2 excitations of the horizon \cite{Grumiller:2016kcp, Ammon:2017vwt}.
}

\acknowledgments

We thank Hamid Afshar, Martin Ammon, St\'ephane Detournay, Hern\'an Gonz\'alez, Alfredo Perez, Stefan Prohazka, Max Riegler, Shahin Sheikh-Jabbari, David Tempo, Ricardo Troncoso, Raphaela Wutte and Hossein Yavartanoo for collaboration on soft hairy aspects and Laura Donnay, Gaston Giribet, Hern\'an Gonz\'alez, Miguel Pino and Andy Strominger for discussions.

DG and WM were supported by the projects of the Austrian Science Fund (FWF), P~27182-N27 and P~28751-N27. During the final stage DG was additionally supported by the FWF project P~30822-N27.
WM is partially supported by the ERC Advanced Grant ``High-Spin-Grav" and by FNRS-Belgium (convention FRFC PDR T.1025.14 and convention IISN 4.4503.15).

\appendix

\section{Algebra conventions}\label{app:A1}

\newcommand{\ia}{a}
\newcommand{\ib}{b}
\newcommand{\ic}{c}
\newcommand{\iT}{0}
\newcommand{\ir}{1}
\newcommand{\ip}{2}
\newcommand{\TJL}{T}

\newcommand{\La}{\TJL^\ia}
\newcommand{\Lb}{\TJL^\ib}
\newcommand{\Lc}{\TJL^\ic}
\newcommand{\Ladn}{\TJL_\ia}
\newcommand{\Lbdn}{\TJL_\ib}
\newcommand{\Lcdn}{\TJL_\ic}
\newcommand{\Lt}{\TJL^\iT}
\newcommand{\Lr}{\TJL^\ir}
\newcommand{\Lp}{\TJL^\ip}
\newcommand{\epsabc}{\epsilon^{\ia\ib\ic}}
\newcommand{\epsabcdn}{\epsilon_{\ia\ib\ic}}
\newcommand{\etaab}{\eta^{\ia\ib}}

We use the $sl(2,\,\mathbb{R})$ generators
\eq{
L_+=\begin{pmatrix*}[r]
0 & 0 \\ -1 & 0 \end{pmatrix*}\qquad\qquad
L_0=\frac12\,\begin{pmatrix*}[r]
1 & 0 \\ 0 & -1
\end{pmatrix*}\qquad\qquad
L_-=\begin{pmatrix*}[r]
0 & 1 \\ 0 & 0
\end{pmatrix*}
}{eq:app1}
which have standard commutation relations
\eq{
[L_+,\,L_-]=2L_0\qquad\qquad[L_\pm,\,L_0]=\pm L_\pm
}{eq:app2}
and the following non-vanishing traces
\eq{
\tr\big(L_+L_-\big)=-1\qquad\qquad\tr\big(L_0L_0\big)=\frac12\,.
}{eq:app3}
For comparison with literature on CS-like theories (see \cite{Merbis:2014vja} and refs.~therein) it is useful to convert into an so$(1,2)$ basis $\La$, given by
\eq{
\Lt = \frac{L_+ + L_-}{2}\qquad\qquad \Lr = \frac{L_+ - L_-}{2}\qquad\qquad \Lp = L_0
}{eq:app4}
with the standard commutation relations ($\ia,\,\ib=\iT,\,\ir,\,\ip$)
\eq{
[\La,\,\Lb]=\epsabc\, \Lcdn\qquad\qquad \epsabcdn = + 1
}{eq:app5}
where indices are raised and lowered by the Minkowski metric 
\eq{
\etaab=2\,\tr\big(\La\Lb\big)=\textrm{diag}(-1,1,1)^{\ia\ib}\,. 
}{eq:app7}
The translation between wedged commutators, index notation and the cross-product notation of \cite{Merbis:2014vja} is then given by
\eq{
[B\stackrel{\wedge}{,}\,A] = [A\stackrel{\wedge}{,}\,B] \equiv \Ladn\,\epsabc\,A_\ib\wedge B_\ic \equiv A \times B = B \times A \,.
}{eq:app6}
Note that in the first and last equality there are two compensating signs. For varying the action \eqref{eq:shw16} the triple product identity 
\eq{
A\wedge[B\stackrel{\wedge}{,}\,C]=B\wedge[C\stackrel{\wedge}{,}\,A]=C\wedge[A\stackrel{\wedge}{,}\,B] 
}{eq:app8}
is useful.

\section{
Chern--Simons-like variables in near horizon expansion}\label{app:A}

Starting with the triad \eqref{eq:bcs} to lowest orders in $r$ we iteratively solve the equations of motion \eqref{eq:EOM}, thereby obtaining the connection from the condition of vanishing torsion and the Schouten one-form from the relation between $f$ and curvature. The final equation of motion then yields conditions for the next subleading term in the triad $e$. In this way solutions to the equations of motion compatible with our boundary and gauge-fixing conditions \eqref{eq:bcs} are found. 

We display below the first couple of terms in such a near horizon expansion. 
\begin{align}
    e_t &= a\,\big(r + \tfrac{1}{6}(3-2\nu^2)\,r^3 + {\cal O}(r^5)\big)\,\frac{L_+ + L_-}{2} \\
    e_\vp &= \gamma\,\big(1+\tfrac12\,\nu^2r^2 + \tfrac{1}{24}\, \nu^2 (7\nu^2-6)\, r^4 + {\cal O}(r^6)\big)\, L_0 \nonumber \\
    &\quad + \big[-\omega\,r + \big(\tfrac34\,\nu(\nu^2-1)\,\gamma - \tfrac16\,(3-2\nu^2)\,\omega\big)\, r^3 + {\cal O}(r^5)\big]\, \frac{L_+ + L_-}{2}  \\
    e_r &= \frac{L_+ - L_-}{2} \displaybreak[1] \\
    \omega_t &= a\,\big(1 + \tfrac12(3-2\nu^2)\,r^2 + {\cal O}(r^4)\big)\,L_0 - a\,\big(\tfrac34\,\nu(\nu^2-1)\,r^3 + {\cal O}(r^5)\big)\, \frac{L_+ + L_-}{2} \\
    \omega_\vp &= \big[- \omega + \tfrac12\big(3\nu (\nu^2-1)\,\gamma - (3-2\nu^2)\,\omega\big)\,r^2 + {\cal O}(r^4)\big]\, L_0  \nonumber\\
    &\quad + \big[\nu^2\,\gamma\,r + \tfrac{1}{12}\nu\big(2\nu(7\nu^2-6)\,\gamma + 9 (\nu^2-1)\,\omega\big)\, r^3 + {\cal O}(r^5) \big]\,\frac{L_+ + L_-}{2} \\
        \omega_r &= \tfrac34\,\nu\big(\nu^2-1\big)\,r^2\,\frac{L_+ - L_-}{2} + {\cal O}(r^4) \displaybreak[1]\\
    f_t &= a\,\big(3\nu(1-\nu^2)\,r^2 + {\cal O}(r^4)\big)\, L_0 + a\,\big(\tfrac12\,(3-2\nu^2)\,r + {\cal O}(r^3) \big)\,\frac{L_+ + L_-}{2} \\
    f_\vp &= \big[\tfrac12(4\nu^2-3)\,\gamma + \big(\tfrac14\,\nu^2\,(16\nu^2-15)\,\gamma + 3\nu (\nu^2-1)\,\omega\big)\,r^2 + {\cal O}(r^4)\big]\, L_0 \nonumber \\
    &\quad + \big[\big(3\nu (\nu^2-1)\,\gamma -\tfrac12\,(3-2\nu^2)\,\omega\big)\, r + {\cal O}(r^3) \big]\,\frac{L_+ + L_-}{2} \\
    f_r &=  \tfrac12\,\big(3-2\nu^2\big)\,\frac{L_+ - L_-}{2} + {\cal O}(r^4)
\end{align}
As in the main text, we assumed above that Rindler acceleration is constant, $a=\rm const.$, so that the near horizon Ward identities \eqref{eq:shw11} imply that the state-dependent functions $\gamma$ and $\omega$ depend solely on the angular coordinate $\vp$. 

The results above together with the formula for the metric 
\eq{
g_{\mu\nu} = 2\,\tr\big(e_\mu \,e_\nu\big)
}{eq:shw17}
yield the line-element \eqref{eq:shw15}.

\section{Warped DGGP boundary conditions}\label{app:A2}

\newcommand{\genvir}{L}
\newcommand{\gencur}{P}

In section \ref{se:3.new} we discussed the transformations $\xi^{\ione}$ that preserve our near horizon warped black hole boundary conditions \eqref{eq:bcs}. An implicit assumption in that derivation (inspired by the Einstein gravity case \cite{Afshar:2016wfy,Afshar:2016kjj}) is that we take the near horizon gauge generator to be independent of the functions specifying the state, $\gamma(\vp)$ and $\omega(\vp)$. This is a subtle, but important difference with respect to the work of Donnay, Giribet, Gonz\'alez and Pino (DGGP) \cite{Donnay:2015abr,Donnay:2016ejv}, where the near horizon Killing vector $\zeta^{\mu}$ is assumed to be independent of the state-dependent function. This approach leads to a different near horizon charges and algebra.

For our near horizon warped black hole boundary conditions in TMG, it is also possible to repeat the analysis of section \ref{se:3too} while assuming the near horizon Killing vector is independent of $\gamma(\vp)$ and $\omega(\vp)$. We may write it in terms of two arbitrary functions $T(\vp)$ and $Y(\vp)$
\eq{
    \zeta = T(\vp) \partial_t + Y(\vp)\partial_\vp + {\cal O}(r^2)\,.
}{eq:DGGPzeta}
By \eqref{eq:diffeos} this implies that the gauge parameters $\xi^\ione$ read
\begin{subequations}
\label{eq:DGGPxi}
\begin{align}
    \xi^e = & \, Y(\vp)\gamma(\vp)  \, L_0 + (a T(\vp) - Y(\vp)\omega(\vp) )\, r\, \frac{L_+ + L_-}{2} + {\cal O}(r^2) \displaybreak[1]\\
    \xi^\omega = &\, (a T(\vp) - Y(\vp)\omega(\vp) )\, L_0 + \nu^2\, Y(\vp)\gamma(\vp) \, r\,  \frac{L_+ + L_-}{2}  + {\cal O}(r^2) \displaybreak[1]\\
    \xi^f =  & \, - \tfrac12(4\nu^2-3)Y(\vp)\gamma(\vp) \,L_0 \\
    & \;\; +  \left( 3\nu (\nu^2-1)Y(\vp)\gamma(\vp) + (\tfrac32-\nu^2)(a T(\vp) - Y(\vp)\omega(\vp) )\,\right)\, r \,\frac{(L_+ + L_-)}{2}  + {\cal O}(r^2) \nonumber
\end{align}
\end{subequations}
which is simply \eqref{eq:shw27} with $\eta(\vp) = (a T(\vp) - Y(\vp)\omega(\vp) )$ and $\eps(\vp) = -  Y(\vp)\gamma(\vp)$. These relations also hold in the variation of the state-dependent functions \eqref{eq:shw28} which now read
\eq{
\delta \gamma(\vp) = \big(Y(\vp) \gamma(\vp) \big)' \, \qquad  \qquad \delta \omega(\vp) = (Y(\vp) \omega(\vp))' - a T'(\vp)\,.
}{eq:DGGPvar}
Here the prime denotes differentiation with respect to $\vp$. The variation of the canonical boundary charges \eqref{eq:shw22} is again integrable and finite. Once integrated the charges give
\eq{
    Q[T,Y] = \frac{1}{8\pi G} \oint \extd \vp \; \Big[ a\, T \, \Big( \gamma + \frac{\omega}{3\nu} \Big) + Y \Big(\gamma\,\omega + \frac{1}{6\nu}\left(\omega^2 + (4\nu^2-3)\gamma^2\right) \Big) \Big]\,.
}{eq:QDGGP}
The Fourier modes $\genvir_n = Q[T=0,Y=e^{in\vp}]$ and $\gencur_n = Q[T=e^{in\vp},Y=0]$ span the `near horizon warped DGGP' algebra, which reads
\eq{
    [\genvir_n,\,\genvir_m] = (n-m)\,\genvir_{n+m} \qquad [\genvir_n,\,\gencur_m] = -m \gencur_{m+n} \qquad [\gencur_n,\,\gencur_m] 
    = \frac{\kappa}{2}\,n\,\delta_{n+m,0} \,.
}{eq:WDGGP}
Like in \cite{Donnay:2015abr}, this is a semidirect sum of the Witt algebra with a $\mathfrak{u}(1)$ current algebra. However, now the $\mathfrak{u}(1)$ current has a non-zero level 
\eq{
\kappa=\frac{a^2}{6\nu G}\,. 
}{eq:whatever}
Note that the level $\kappa$ is determined entirely by the gravitational
anomaly and thus vanishes only in the limit of infinite $\nu$. As observed in \cite{Donnay:2015abr} for the BTZ black hole, also the warped black hole entropy \eqref{eq:SALPSS} is given by $\gencur_0$ times the inverse temperature $\beta = 2\pi/a$. The warped conformal analysis done in AdS \cite{Afshar:2016wfy} can be repeated verbatim and leads to the same result for entropy. Linearity of the entropy in $P_0$ is also predicted by the warped conformal generalization of the Cardy formula for vanishing central charge $c=0$, see Eq.~(49) in \cite{Detournay:2012pc}.

\section{Entropy from a boundary term at the horizon}\label{app:B}

In this appendix we generalize the arguments from \cite{Bunster:2014mua} for computing the entropy of black holes to Chern--Simons-like theories of gravity. Obviously, for the results here to hold, we would need a suitable CS-like theory of gravity that allows for non-trivial and non-degenerate solutions whose metric interpretation is that of a stationary black hole with regular horizon at a radial coordinate $r=r_+$ and inverse temperature $\beta = 2\pi/a$. We will assume here that such solutions exist in the theory of our interest, but other than that the derivation holds for general CS-like theories.

We start with the action \eqref{eq:shw20} and add a boundary term to ensure that the variational principle is well-defined
\eq{
    I = \frac{k}{2\pi} \int \tr \Big(g_{\ione\itwo}\, a^\ione \wedge\extd a^\itwo + \frac13\, f_{\ione\itwo\ithr}\, a^\ione\wedge a^\itwo \wedge a^\ithr \Big) + B_{r_+}
}{actionwbt}
The Hamiltonian form of the action is obtained by performing a space-time split 
\begin{align}
	a^\ione = a_t{}^\ione\extd t+a_i{}^\ione \extd x^i\,,
\end{align}
where $i$ represents spatial indices. This leads to the Lagrangian density
\begin{align}
	\mathcal{L}=\tr \left(-\epsilon^{ij}g_{\ione\itwo}a_i{}^\ione \partial_t a_j{}^\itwo + 2 a_t{}^\ione \phi_\ione \right)\,,
	\label{lagrangiandensity}
\end{align}
with the constraint functions $\phi_\ione$.
Following \cite{Bunster:2014mua}, we will pass to Euclidean signature via a Wick rotation $t=-i\tau$. Now, both $\vp$ and $\tau$ are periodic with periodicities $\vp \sim \vp+2\pi$ and $\tau \sim \tau+\beta$. 

The entropy of a black hole can be obtained by evaluating the action on-shell, i.e., on the black hole solution. The action has to be of such a form that if we demand it to be stationary with some boundary conditions at infinity, the equations of motion should hold everywhere. 
We choose to work in the Hamiltonian form because black hole solutions are time independent as they describe a thermodynamic system in equilibrium. Then, the first term in \eqref{lagrangiandensity} vanishes and the constraint $\phi_\ione=0$ has to hold on-shell. 
The entropy comes from a contribution to the action at the horizon $r_+$ that stems from demanding a regular solution at $r_+$.

The variation of the action \eqref{actionwbt} reads on-shell
\begin{align}
	\delta I\big|_{\textrm{\tiny EOM}} = \delta B_{r_+} - i \frac{k}{\pi}\, \int_{r=r_+} \extd \tau \extd\vp \,\tr\big(g_{\ione\itwo} a_\tau{}^\ione \delta a_\vp{}^\itwo\big)\,.
\end{align}
From our assumptions on the stationarity of the solution it follows that we can readily evaluate the time integral as
\begin{align}\label{eq:delB}
	\delta I\big|_{\textrm{\tiny EOM}} = \delta B_{r_+} - i\frac{k}{\pi}\beta \oint_{r=r_+} \extd\vp \, \tr\left(g_{\ione\itwo} a_\tau{}^\ione \delta a_\varphi{}^\itwo \right) \,.
\end{align}
Now, $B_{r_+}$ is chosen such that $\delta I=0$. This requires us to integrate the surface term in the above expression, which may in general be non-trivial. Fortunately, as explained in section \ref{se:cs3}, by consistency of the boundary conditions, the time component of the fields $a_t{}^\ione$ have to be proportional to a boundary condition preserving gauge transformation of $a_{\vp}{}^\ione$. This implies that the condition that the surface term in \eqref{eq:delB} is integrable is equivalent to the condition that the canonical boundary charges \eqref{eq:shw22} are integrable at the horizon. Provided that these conditions are met, and the gauge parameters do not depend on the charges (which is the case for our near horizon boundary conditions) we can readily integrate \eqref{eq:delB} and write on-shell
\begin{align}
	B_{r_+} = i\frac{k}{\pi}\beta \oint_{r=r_+} \extd\vp \, \tr\left(g_{\ione\itwo} a_\tau{}^\ione a_\varphi{}^\itwo \right)\,.
\end{align}
When \eqref{actionwbt} is evaluated on-shell for stationary configurations, the canonical bulk action vanishes, and the entropy is given by the contribution of the boundary term at the horizon. Going back to Lorentzian signature, this yields
\begin{align}
	S = -\frac{k}{\pi}\beta \oint_{r=r_+} \extd\vp \, \tr\left(g_{\ione\itwo} a_t{}^\ione a_\varphi{}^\itwo \right)\, \,. \label{genentr}
\end{align}
One can now easily verify that the TMG field space metric \eqref{eq:shw24}, together with the near horizon solution for the fields $a^\ione$ given in appendix \ref{app:A}, reproduce the entropy \eqref{eq:entropy1} at $r=0$.

\addcontentsline{toc}{section}{References}

\providecommand{\href}[2]{#2}\begingroup\raggedright\endgroup

\end{document}